\documentclass[journal=jacsat,manuscript=article,layout=twocolumn]{achemso}
\usepackage{graphicx}
\usepackage{caption}
\usepackage{subcaption}
\usepackage{comment}
\usepackage{xcolor}
\usepackage{multirow,makecell,tabularx,bigstrut}
\usepackage[version=3]{mhchem} 

\usepackage{xr}
\usepackage[pdftex]{hyperref}
\usepackage{hypernat}
\hypersetup{colorlinks,%
	citecolor=black,%
	anchorcolor=black,%
	menucolor=black,%
	runcolor=black,%
	filecolor=black,%
	linkcolor=black,%
	urlcolor=black}

\let\oldmaketitle\maketitle
\let\maketitle\relax

\author{Francesco Maria Bellussi}
\affiliation[Politecnico di Torino]{Department of Energy, Politecnico di Torino, Torino, Italy}
\author{Matteo Ricci}
\affiliation[MaterialX LTD]{MaterialX LTD, Bristol, UK}
\author{Matteo Fasano}
\email{matteo.fasano@polito.it}
\affiliation[Politecnico di Torino]{Department of Energy, Politecnico di Torino, Torino, Italy}
\email{matteo.fasano@polito.it}
\phone{}
\author{Otello Maria Roscioni}
\affiliation[MaterialX LTD]{MaterialX LTD, Bristol, UK}
\email{om.roscioni@materialx.co.uk}
\phone{}

\title[]{Mesoscopic modelling of bio-compatible PLGA polymers with coarse-grained molecular dynamics simulations}


\begin{document}

\begin{tocentry}
\includegraphics[width=.9\textwidth]{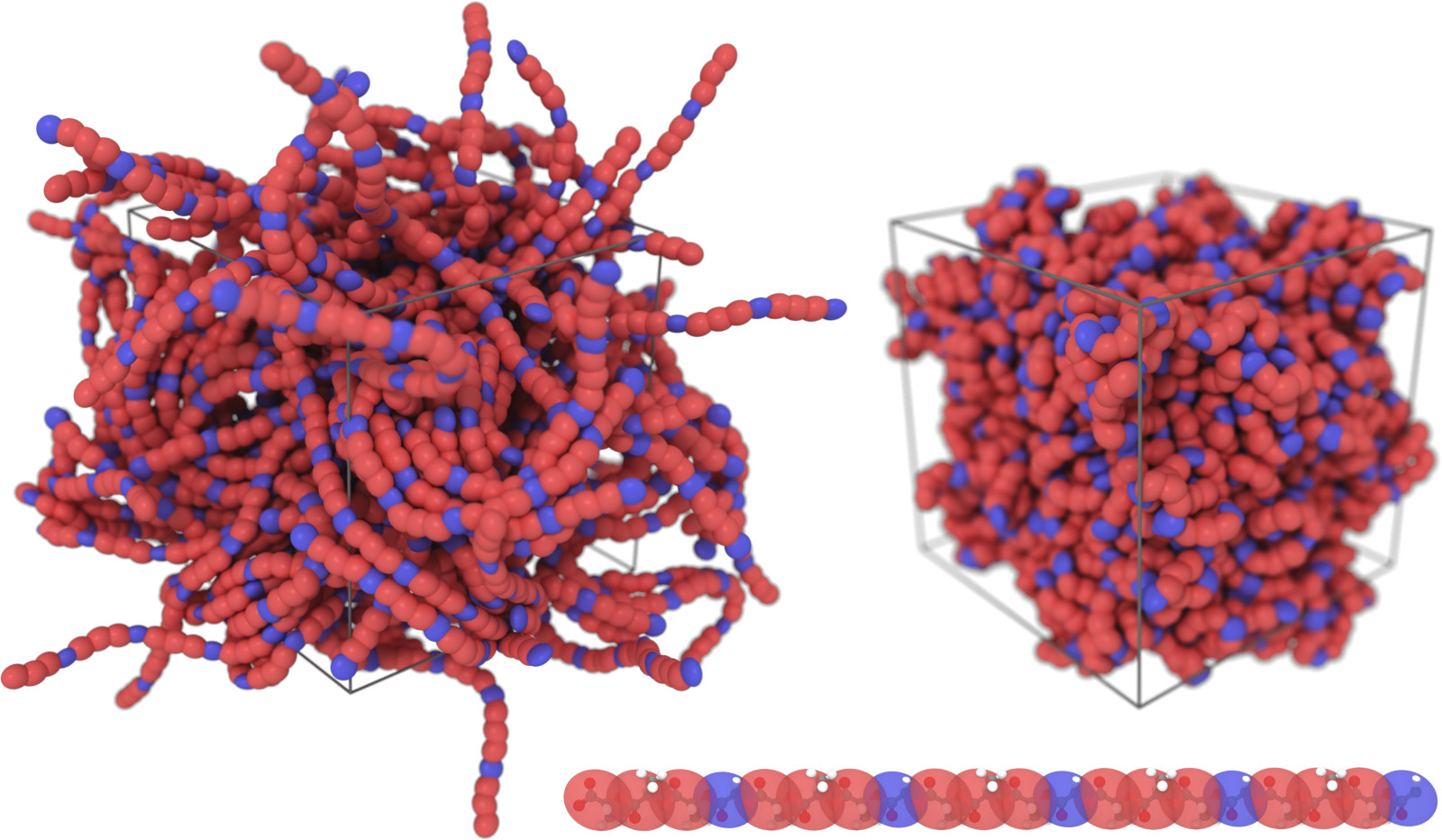}
\label{fig:TOC Graphic}
\end{tocentry}
%
\twocolumn[
    \begin{@twocolumnfalse}
    	\oldmaketitle
        \begin{abstract}
           A challenging topic in materials engineering is the development of numerical models that can accurately predict material properties with atomistic accuracy, matching the scale and level of detail achieved by experiments.
           In this regard, coarse-grained (CG) molecular dynamics (MD) simulations are a popular method for achieving this goal. Despite the efforts of the scientific community, a reliable CG model with quasi-atomistic accuracy has not yet been proposed for the design and prototyping of materials, especially polymers. In this paper we describe a CG model for polymers, focusing on the bio-compatible poly(lactic-co-glycolic acid) (PLGA), based on a general parametrization strategy with a potentially broader field of applications. In this model,
           polymers are represented with finite-size ellipsoids, short-range interactions are accounted for with the generalized Gay-Berne potential, while electrostatic and long-range interactions are accounted for with point charges within the ellipsoids. The model was validated against its atomistic counterpart, obtained through a back-mapping process, by comparing physical properties such as glass transition temperature, thermal conductivity, and elastic moduli. We observed quantitative agreement between the atomistic and CG representations, thus opening up the possibility of adopting the proposed model to expand the domain size of typical MD simulations to dimensions comparable to those of experimental setups.
        \end{abstract}
    \end{@twocolumnfalse}
]

\section{Introduction}
The theoretical modelling of bio-compatible polymers is an interesting technical challenge due to the emergent nature of their properties \cite{asadi2020common,puertas2021emerging,terzopoulou2022biocompatible}. At the most basic level, the physical properties of polymers depend on their chemical composition and molecular structure \cite{de2024diffusion}. At the mesoscopic level, their properties also depend on the supramolecular arrangement and entanglement of polymeric chains, the presence and size of ordered/crystalline domains, the relaxation dynamics, and the presence of structural defects or cross-linking in the condensed structure. An obvious way to capture these aspects in computer simulations is to use large domains to represent the sample and simplified models to represent the material.

Due to the size of the systems involved, the less relevant degrees of freedom, e.g. the high-frequency motion of light atoms, are typically omitted from the simulations \cite{dhamankar2021chemically}. This procedure is known as \emph{coarse-graining} and involves grouping atoms into larger units called beads. 
This abstraction focuses on capturing the essential physical properties and interactions of polymers \cite{muhammad2023atomistic,muhammad2023mesoscopic}, allowing computer simulations complex enough to capture phenomena spanning over different spatial and temporal scales. Several coarse-grained (CG) models of polymers have been proposed during the last two decades \cite{lee2009coarse,noid2013perspective,kempfer2019development,joshi2021review,ye2021machine,alessandri2021martini}, each following different approaches for the mapping of atoms into beads. A way to mitigate the information lost in the mapping process is to use finite-size aspherical particles as CG units \cite{molc,cohen2021anisotropic,shen2014anisotropic}. 

In particular, the CG model named MOLC \cite{molc} has been developed to represent liquids \cite{bellussi2021} and organic functional materials \cite{molc,roscioni2023osc} with atomistic-like accuracy. In the MOLC model, molecular functional groups (e.g. aromatic moieties) are mapped to ellipsoidal beads. The pair potential acting between ellipsoids is a sum of the Gay-Berne potential \cite{berardi1995generalized,berardi1998gay} and classic electrostatics, accounted for with point charges embedded in the parent ellipsoidal bead. In large molecules, ellipsoids are bonded with a two-body potential that simultaneously controls the position and orientation of the bonded pair, making use of the finite-size nature of the ellipsoids. The resulting CG model reproduces accurately the excluded volume of the corresponding atomistic model, yielding precise predictions of the condensed-phase properties and enabling a 1-to-1 mapping of the CG trajectory back to the corresponding all-atom (AA) representation. Consequently, the MOLC model enables a seamless transition between CG and AA representations, depending on the desired level of simulation detail \cite{roscioni2023osc}.
The MOLC model is implemented in a custom version \cite{user-molc} of the molecular dynamics (MD) software LAMMPS \cite{thompson2022lammps,nguyen2019aspherical}, while the force field parameters and molecular topologies are stored in the LAMMPS-template (LT) format and processed with the software Moltemplate \cite{moltemplate}.

This work aims to develop and validate a structurally accurate CG model for the bio-compatible poly(lactic-co-glycolic acid) (PLGA) polymer. The focus here is to compute the mechanic and thermodynamic properties of PLGA from CG MD simulations and to assess the effect of bulk morphology on the resulting functionality. More specifically, we compare a semi-crystalline sample of PLGA with an amorphous sample. The two samples are generated at the CG level and are subsequently back-mapped to the AA level to obtain reference data for validation purposes.

The article is structured as follows. In the Computational Methods, we present the parametrization of the PLGA CG model and the corresponding back-mapping procedure to the atomistic representation. Then, we describe the simulation methods to assess the thermal conductivity and elastic properties of both the CG and back-mapped AA models. In the Results and Discussion sections, we compare the physical observables obtained from AA and CG simulations. Finally, we present concluding remarks and discuss future research prospects in this field.

\section{Computational methods}

\subsection{Parametrization}
\begin{figure*}[t!]
     \centering
     \includegraphics[width=0.8\textwidth]{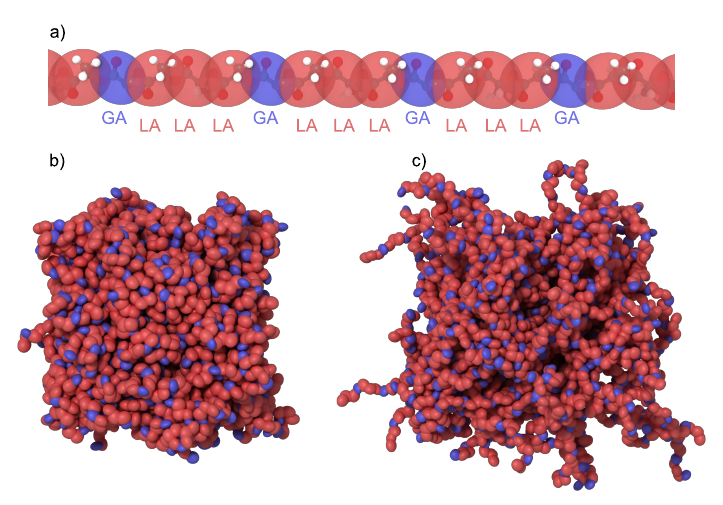}
     \caption[CG samples of PLGA]{(a) Mapping of the CG model of a PLGA chain to its AA representation. LA (red beads) and GA (blue beads) indicate the lactic acid and glycolic acid monomers, respectively. Each monomer is described by a single CG bead. CG models of (b) amorphous and (c) semi-crystalline PLGA samples.}
     \label{PLGA-samples}
\end{figure*}
The MOLC model is based on finite-sized ellipsoids which have a moment of inertia, rotational energy, and angular momentum. Each ellipsoidal bead replaces a molecular fragment, thus reducing the total degrees of freedom and allowing the use of a simulation time-step ten or twenty times higher than that of a classical molecular dynamics simulation \cite{bellussi2021}. Here, each lactic acid (LA) and glycolic acid (GA) monomer is represented with a single ellipsoid in the mesoscopic model, thus reducing the number of particles in the system by a factor of 8.25 compared to the atomistic model.

The bonded and non-bonded potentials for LA and GA monomers are parameterized following a bottom-up approach \cite{molc}, considering reference data from AA MD simulations carried out with the OPLS-AA/M force field \cite{oplsaam}. The polymerization of PLGA is obtained via the condensation of the alcoholic and carboxylic functional groups of the GA and LA monomers, which proceeds with the elimination of one water molecule per monomer added. From a modelling perspective, the \emph{effective} repeating units are thus the molecular fragments defined by breaking the sigma bond between the ester oxygen and the alkyl carbon atoms along the polymeric chain. 
The repeating LA and GA units carved from the polymer chain (see Fig. S1 in the supporting information) are capped by adding hydrogen atoms on dangling bonds. The following structures are thus used to parameterize the CG beads: acetic acid for the GA; propionic acid (for the electrostatics) and isobutyric acid (for the Gay-Berne ellipsoids) for the LA. The latter choice for the LA bead yields an ellipsoid symmetric to the polymeric chain while keeping an asymmetric electrostatic potential that maps to the corresponding atomistic model. Additional details on the reference atomic charges used to fit the CG molecular electrostatic potential are shown in the supporting information, Fig. S2. An overview of the overlapping AA and CG representations of a PLGA oligomer is shown in Fig. \ref{PLGA-samples}a.

A directional bond potential for the PLGA backbone was modelled on reference AA-MD simulations of LA-LA, LA-GA, and GA-LA dimers in the gas phase. Potential energy curves were obtained from Boltzmann inversion of the probability distribution histograms of the scalar products between the fragments' axes of inertia. The calculation of the axes of inertia is the (forward) mapping operation that transforms the AA coordinates into CG ones. The three series of bond potentials were sufficiently similar to each other (see supporting information, Fig. S3) to allow the use of only one type of bond potential for the backbone, regardless of the type of bonded monomers.

When the MOLC model is used to create a CG force field for macromolecules and polymers, a large number of ellipsoids end up connected with directional bonds and overlapping in space. This arrangement is obtained by zeroing the pairwise interactions between bonded pairs, as in atomistic type-I force fields. The intended consequence of this choice is that the MOLC model accurately reproduces the molecular excluded volume of the corresponding atomistic fragments, yielding realistic solid-state properties such as density and radial distribution functions \cite{molc,roscioni2023osc}.
The unintended consequence is that the model leads to a tighter packing of ellipsoids in space, resulting in an overestimate of the intermolecular energy. To compensate for this systematic error, the well-depth parameter $\varepsilon_0$ and ellipsoid's axes $\sigma_i$ of the Gay-Berne potential are scaled to reproduce the density and cohesion energy computed from reference AA-MD simulations obtained by back-mapping the CG samples back to the corresponding AA structure.
This parametrization strategy has been recently validated on linear alkanes from 2 to 16 carbon atoms \cite{roscioni2023_molc_parametrization}, and it is further tested in this work for predicting the structural and mechanical properties of PLGA polymers.

More specifically, an amorphous sample of pristine poly(lactic acid) (PLA) polymer was created with CG-MD simulations and back-mapped to its corresponding AA representation with the open-source program Backmap \cite{backmap}. The cohesion energy of models with a long-range electrostatic solver was computed with custom commands called \verb|compute inter| and \verb|compute inter/molc| for AA and CG models, respectively. The code of these commands is largely based on the existing \verb|compute group/group| framework, with a different correction term for the intra-molecular electrostatic interactions \cite{user-molc,roscioni2023_molc_parametrization}.
A scaling factor of 0.7 was obtained for the $\varepsilon_0$ of PLA. The same scaling factor was used for the $\varepsilon_0$ of the GA repeating unit. The optimised CG force field for PLGA is provided as supporting information in the LT format.

MD simulations of the CG samples used a 10 fs timestep for the production run, even if a 20 fs timestep could be used in conjunction with a thermostat. A cut-off radius of 14 \AA{} was considered for both short-distance Gay-Berne and long-distance electrostatic interactions, and a skin distance of 4 \AA{} was used to build the neighbour list.
The back-mapped AA samples were first relaxed in the NPT ensemble at 298 K, 1 atm, with a timestep of 1 fs until the total energy reached equilibrium, which typically required about 2 ns. The van der Waals and electrostatic interactions were modelled with a cut-off radius of 12 \AA, while the long-range electrostatic interactions were calculated using the particle-particle particle-mesh (PPPM) method \cite{hockney2021computer}.
All MD simulations for CG and AA samples were performed with the latest stable version of LAMMPS (2 Aug 2023) \cite{LAMMPS,plimpton1995}, modified to include the MOLC package \cite{user-molc}.

\subsection{Sample Preparation}

Two computational samples of PLGA with different morphologies were created by varying the simulation settings. Both samples contain 90 PLGA chains with a 75:25 LA to GA ratio. Each PLGA chain has 12 repetitions of a fixed LA-LA-LA-GA sequence, for a total of 48 monomers for each chain.

The initial configuration of the CG samples was created using Moltemplate \cite{moltemplate} to place PLGA chains on a sparse cubic grid, with each chain randomly rotated around its centre of mass. Then, a compression and relaxation protocol already used to create atomistic samples of PLGA \cite{bellussi2022wettability} was used to obtain CG samples with a realistic initial density. Two polymer morphologies, namely amorphous and semi-crystalline (see Figs. \ref{PLGA-samples}b--c), were created by tweaking the duration of the compression and relaxation stages of the samples and the force field parameters.

The sample of semi-crystalline PLGA was obtained by compressing an initial configuration where the molecules were placed on the nodes of a cubic grid with spacing 185.4 \AA{} with random orientation. Periodic boundary conditions were employed to ensure continuous packing throughout space. The initial configuration was compressed at a constant rate for 0.49 ns to a target density of 1.16 g/cm$^3$, keeping the temperature at 1 K using a Langevin thermostat with a coupling constant of 10 ps and a timestep of 10 fs. After the isotropic compression, the sample was heated to 300 K at a rate of 0.75 K/ps and further equilibrated for 2 ns in the canonical ensemble (NVT) with a timestep of 20 fs. A custom bond potential with harmonic constants of 20 kcal/mol was used during the compression phase to keep the chains in the extended configuration. The intermolecular interactions were also modified by including only the Gay-Berne part of the pair potential. The removal of electrostatic interactions approximates the presence of a solvent, screening the molecular electrostatic potential and lubricating chain motions during the compression stage. The sample finally equilibrated in the NPT ensemble using the regular CG potential and a timestep of 20 fs, yielding an almost cubic sample with a side of approximately 71 \AA.

The sample of amorphous PLGA was obtained with a protocol similar to that of the semi-crystalline sample but with a 10x compression rate. During the compression, a modified CG force field was considered, in which the pair potential contains only short-range interactions without long-range electrostatics, as for the semi-crystalline sample. At all stages of this protocol, the bond potential was the one fitted to the reference AA simulations, which allows the chains to be fully flexible. The resulting sample is characterised by virtually no inter-digitation between the chains, which are rolled up into blobs. The sample was compressed by keeping the temperature at 1 K until a target density of 1.16 g/cm$^3$ was reached. The sample was then heated to 300 K in the canonical ensemble and equilibrated in the NPT ensemble with a timestep of 20 fs, yielding an almost cubic sample with a side of approximately 71 \AA.

\subsection{Thermal conductivity}
\label{methods_thermal}

The thermal conductivity of the PLGA CG and back-mapped AA systems was evaluated with the non-equilibrium molecular dynamics (NEMD) approach \cite{wirnsberger2015enhanced}. This method is implemented by imposing a thermal gradient on the system and calculating the subsequent heat flux. The thermal gradient is imposed by dividing the simulation box into an odd number of bins (see Fig. S4a in the supporting information). The central bin corresponds to the hot region, thermalized at 320 K, while the two terminal bins, each half the size of a regular one, are thermalized at 280 K using a Langevin thermostat. To ensure a comparable statistical sampling, the CG samples were divided into 13 bins, while the AA samples into 17 bins.
The equations of motion are integrated with the NVE ensemble to allow the heat flux to flow from the central to the terminal bins. The thermal conductivity is then computed through the Fourier model:
\begin{equation}
  \mathbf{q} = \lambda_c \nabla T ,
  \label{eqn:thermal_conductivity}
\end{equation}
where $\mathbf{q}$ is the specific heat flux, $\lambda_c$ is the thermal conductivity, and $\nabla T$ is the temperature gradient. $\nabla T$ is evaluated and averaged over the middle bins to avoid any possible artefact arising from the thermostat. First, we performed 1 ns simulations to reach the non-equilibrium steady state. Then we acquired data related to the thermal gradient and heat flux on a trajectory of 2 ns, using the last 0.5 ns of the production run.

\begin{figure*}[t!]
	\centering
	\includegraphics[width=\textwidth]{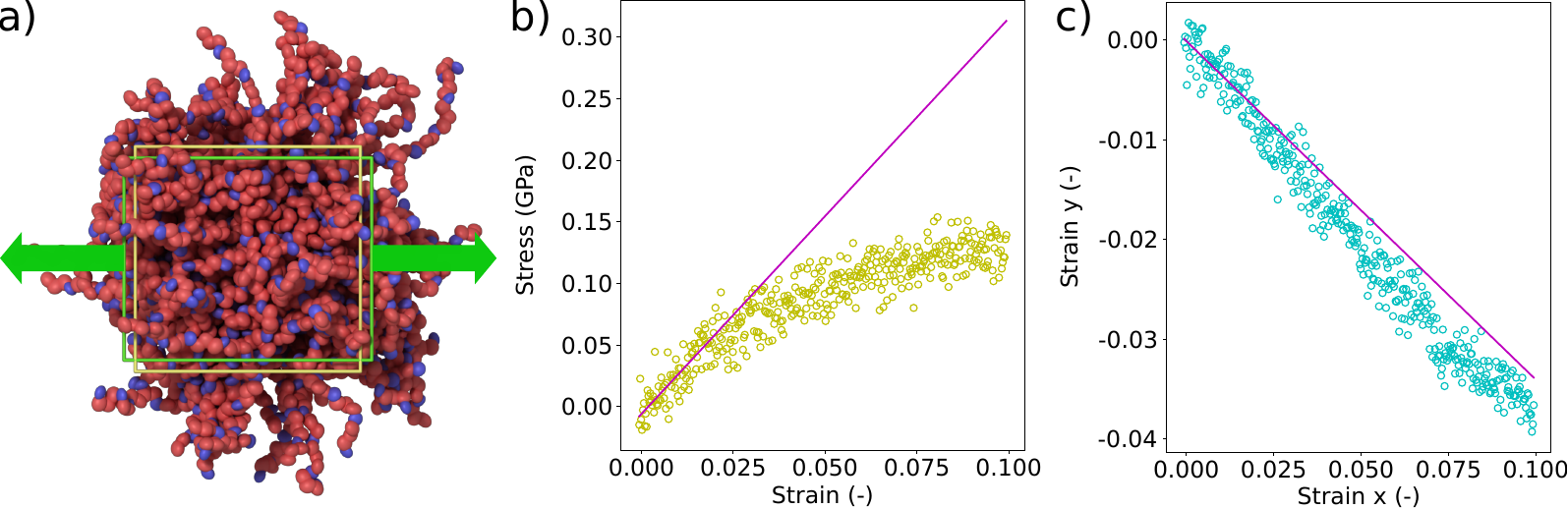}
	\caption[]{(a) The green arrows indicate the deformation applied to the simulation box in stress--strain simulations. Yellow and green lines indicate the initial and deformed box, respectively. The deformed box shows both the elongation along the applied deformation direction and the shrinkage in the transversal direction. (b) Example of a stress$_i$--strain$_i$ curve along the $i$ direction. (c) Example of the strain$_{j,k}$--strain$_i$ curve. The violet lines show the linear extrapolation between 0 and 0.025 of applied strain.}
	\label{stress-strain-curves}
\end{figure*}

For CG simulations, we implemented a new LAMMPS command to compute the average temperature in each bin associated with all the degrees of freedom (rotational + translational) \cite{user-molc}. Despite the recent increasing interest in rigid body dynamics, which is testified by the large number of new packages developed in LAMMPS \cite{nguyen2019aspherical}, a command that can evaluate the average temperature per bin associated with the roto-translational kinetic energy is not present in the official LAMMPS distribution. Consequently, we defined a new compute command called \verb|compute erotate/asphere/atom| to access the rotational kinetic energy using the powerful \verb|atom/chunk| framework used in LAMMPS to partition per-atom quantities into sets called chunks. The rotational energy computed in this way is added to the translational kinetic energy, which is evaluated with the standard LAMMPS command \verb|compute ke/atom|. The roto-translational temperature is then computed with the following equation:
\begin{equation}
\label{eqn:thermal_conductivity2}
\begin{aligned}
  KE &= \sum_{i=1}^{n} \frac{1}{2}m_iv_i^2 + \frac{1}{2}I_i\omega_i^2 \\ 
  &= \frac{DOF}{2}k_B T
\end{aligned}
\end{equation}
where $KE$ is the total kinetic energy per chunk, $i$ is the $i$-th atom in the chunk, $m$ and $v$ are its mass and translational velocity, respectively, $I$ and $\omega$ are its moment of inertia and rotational velocity, respectively, $DOF$ is the number of degrees of freedom (6 in 3D), $k_B$ is the Boltzmann constant, and $T$ is the roto-translational temperature of the chunk. The thermal conductivity is evaluated along the three orthogonal directions to compensate for possible inhomogeneities in the structure and to increase the statistics. The LAMMPS scripts for computing thermal conductivity are provided in the supporting information.

\subsection{Elastic properties}

The elastic properties of the CG and the corresponding back-mapped systems are evaluated through NEMD. In this case, after the relaxation process, we applied a box deformation along one orthogonal direction ($i$) at a constant speed of 7.3 \AA/ns, while leaving the other two vectors ($j, k$) free to relax (see Fig. \ref{stress-strain-curves}a). This process is implemented by applying a thermostat at the equilibrium temperature (298 K) and a barostat at 0 atm in the two vectors orthogonal to the deformation. This setup allows the calculation of not only the stress--strain curve and the Young's modulus ($E$), but also of the transverse strain associated with the shrinkage on the two directions orthogonal to the applied deformation ($\epsilon_{j,k}-\epsilon_i$), and thus the Poisson ratio ($\nu$). The elastic modulus is evaluated considering the slope of the linear regression of the initial portion of the stress--strain curve (between 0 and 0.025 strain, see Fig. \ref{stress-strain-curves}b). Similarly, the Poisson ratio is evaluated considering the linear regression slope of the initial portion of the strain-strain curve (between 0 and 0.025 strain, see Fig. \ref{stress-strain-curves}c). Finally, under the assumption of an isotropic material, the transverse elastic modulus ($G$) is evaluated as:
\begin{equation}
\label{eqn:thermal_conductivity3}
  G = \frac{E}{2(1+\nu)}.
\end{equation}
The LAMMPS scripts for computing the elastic properties are provided in the supporting information.

\section{Results and Discussions}

The main goal of this work is to evaluate the MOLC model of PLGA polymers by studying two samples with an amorphous and semi-crystalline morphology. The evaluation of the MOLC model is carried out by computing bulk properties such as density, elastic moduli (Young's Modulus $E$, Poisson Ratio $\nu$, and Tangential Modulus $G$), and the thermal conductivity $\lambda_c$ obtained from CG simulations and comparing them with the corresponding properties obtained from back-mapped AA samples. This protocol ensures that the AA simulations inherit the morphology of their parent CG samples, thus allowing us to attribute any discrepancy between the CG and AA samples only to the different granularity of the two models.

The PLGA samples obtained after the initial compression and relaxation stages were relaxed in the isobaric-isothermal ensemble (NPT) for 20 ns using the PLGA CG force field. The equilibrium density of the amorphous and semi-crystalline CG samples was 1.256(2) g/cm$^3$ and 1.248(2) g/cm$^3$ respectively, in excellent agreement with the value of the back-mapped AA models: 1.280(2) g/cm$^3$ and 1.288(2) g/cm$^3$ respectively, corresponding to an error of 3\%. The density of both CG and AA models is also in excellent agreement with the experimental value of 1.30 g/cm$^3$ \cite{plga_prod,kapoor2015plga}.

\begin{figure}[t!]
	\centering
	\includegraphics[width=0.5\textwidth]{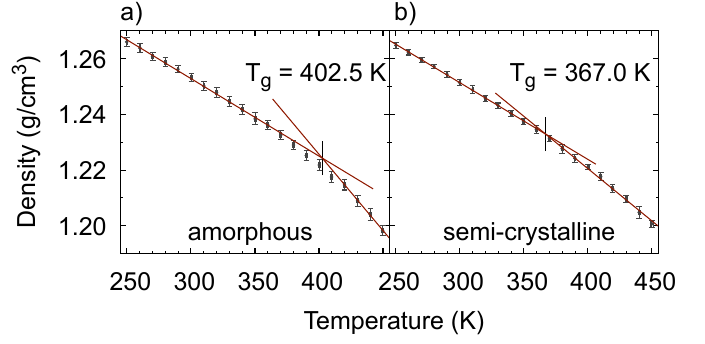}
	\caption[]{Glass transition temperature $T_g$ of (a) amorphous and (b) semi-crystalline CG PLGA samples. The $T_g$ is obtained from the intersection of the linear extrapolation in the plastic and glassy states.}
	\label{Tg_cg}
\end{figure}

The glass transition temperature ($T_g$) of the two CG samples was computed by performing a temperature scan between 250 K and 450 K ($\Delta T = 10 K$) and monitoring the density variation \cite{plga9}. The temperature was changed with a linear ramp lasting 2 ns, followed by an equilibration run of 6 ns and a production run of 1 ns, using a timestep of 10 fs. Thanks to the higher computational efficiency of CG simulations, we carried out five sequential heating-up and cooling-down cycles. This approach aims to improve statistical accuracy, reduce computational noise, and mitigate hysteresis effects. In Figs. \ref{Tg_cg}a--b, we report the density of PLGA as a function of temperature for the two CG models, identifying the two branches of the plastic and glassy states. These branches are fitted with two linear regressions, whose interception point gives $T_g$. Due to the high computational cost of evaluating the $T_g$ for the back-mapped AA samples, we limit the comparison to the $T_g$ of 328 K of an amorphous AA PLGA sample obtained with a single heating-up cycle \cite{bellussi2022wettability} and the experimental $T_g$, which falls in the range of 320--330 K \cite{plga_prod,kapoor2015plga}. The $T_g$ obtained from CG MD simulations is 402.5 K for the amorphous phase and 367.0 K for the semi-crystalline phase.
The difference between the glass transition temperature in the two phases shows that the CG model is detailed enough to be affected by the morphology of PLGA. The observed difference between the $T_g$ computed at the CG and AA levels (40 to 70 K) is unlikely to depend on the thermal history of the AA sample, as its morphology is consistent with that of the back-mapped amorphous sample, as shown in Fig. S6 in the supporting information. Since the intermolecular interactions in the CG force field were optimized to reproduce the reference value from AA samples, we attribute this discrepancy to the reduction of the inter-chain degrees of freedom in the PLGA backbone, as it is well-established that torsional potentials are strongly correlated to the predicted glass transition of linear polymers \cite{canales2007,jin2022}.

\begin{figure*}[t!]
     \centering
     \includegraphics[width=.85\textwidth]{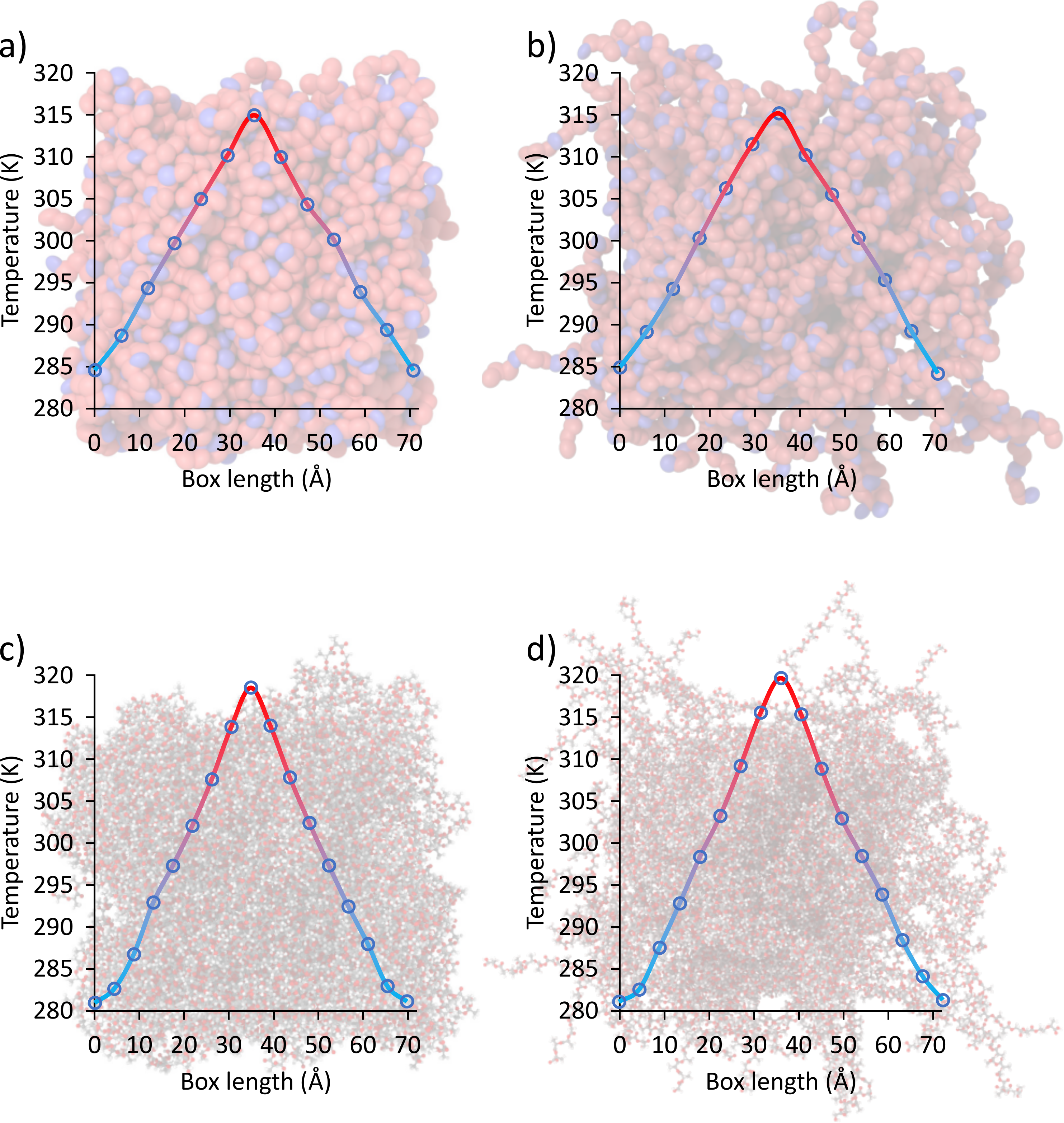}
     \caption[]{Temperature profile of: CG samples with (a) amorphous and (b) semi-crystalline morphology; 
     AA samples with (c) amorphous and (d) semi-crystalline morphology. All samples have an approximately cubic cell with a side length of 71 \AA.}
     \label{temp-profiles}
\end{figure*}

The thermal conductivity was computed on the CG samples following the protocol described in the computational methods section. The temperature profile for the CG samples in Fig. \ref{temp-profiles}a-b shows that the measured temperature in the hot and cold bins is approximately 5 K off the target value. We attribute this discrepancy to the small number of CG beads present in the bins (see Figure \ref{binning}b-c), which makes the temperature an ill-defined quantity. This temperature discrepancy can be eliminated by decreasing the thermostat damping parameter from 1000 to 100 ps, which also increases the cumulative energy exchanged with the thermostat, which does not affect the calculated thermal capacity. The temperature discrepancy can also be eliminated using a $2\times2\times2$ super-cell, resulting in a number of particles per bin comparable to that of the AA samples.
Indeed, no temperature discrepancy is observed in the thermalized bins for the AA samples (see Figs. \ref{temp-profiles}c--d), which contain an average of 2300 atoms per bin.

\begin{figure*}[t!]
	\centering
	\includegraphics[width=\textwidth]{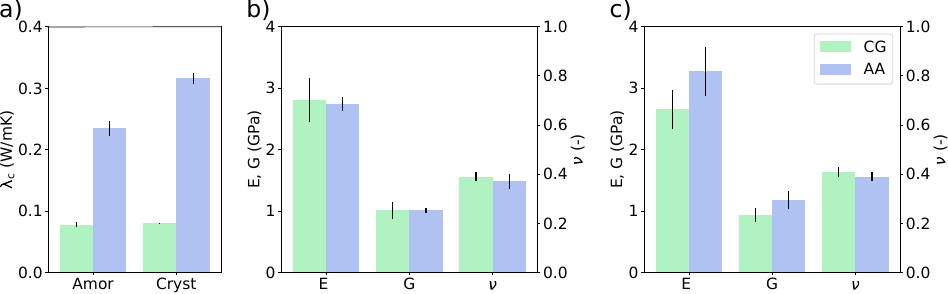}
	\caption[]{(a) Comparison between the thermal conductivity computed from CG (green) and AA (blue) models for both the amorphous and semi-crystalline morphologies. Comparison between the elastic ($E$) and shear ($G$) moduli and Poisson ratio ($\nu$) computed from CG and AA models in the (b) amorphous and (c) semi-crystalline morphologies.}
	\label{bulk-properties}
\end{figure*}

The thermal conductivity obtained from CG and AA simulations is shown in Fig. \ref{bulk-properties}a. The thermal conductivity of AA samples agrees well with experimental findings and previous numerical simulations of polymers and was found to be 
0.23(1) W/mK for the amorphous sample and 0.32(1) W/mK for the semi-crystalline one. For the CG models, the thermal conductivity is around 0.08(1) W/mK, regardless of the sample morphology. To the best of our knowledge, experimental data for the thermal conductivity of PLGA are not available in the literature. Nevertheless, our results are comparable to those reported for poly-lactic acid (PLA), which are in the range of 0.18--0.20 W/mK \cite{lebedev2017poly,guo2019electrical}.

The thermal conductivity obtained from CG simulations is consistently lower than the one from the AA counterparts. Differently from what was observed for the $T_g$, the CG model of PLGA does not yield a different thermal conductivity $\lambda_c$ for the amorphous and semi-crystalline phases. Similarly to previous works \cite{muhammad2023mesoscopic}, the lower thermal conductivity can be attributed to the loss of degrees of freedom and to the suppression of high-frequency vibrational modes in the CG model, which also reduces the sensitivity of the CG model to the sample morphology.

The elastic properties computed on amorphous and semi-crystalline samples with CG and AA models (Figs. \ref{bulk-properties}b--c) agree well with experimental data available in the literature \cite{damadzadeh2010effect}. For the amorphous sample, the difference between the AA and CG models is within the error bars for both Young's modulus and Poisson's ratio, and hence for the tangential elastic modulus. The equivalence of Young's modulus and Poisson's ratio indicates the same stress and lateral shrinkage response between the AA and CG models for lateral and longitudinal deformations. For the semi-crystalline sample, the Young's modulus computed with the CG model is smaller than the corresponding AA one. In contrast, the Poisson ratio exhibits the same lateral shrinkage response to longitudinal deformation for both the AA and CG models. The derivative transverse modulus in the CG sample is consistently lower than that of the AA sample, even though it is still within the error bar.

As seen for thermal conductivity, the elastic properties computed at the CG level are not sensitive to the morphology of the simulated sample (for a direct comparison, see Fig. S5 in the supporting information).
In contrast, the same properties computed at the AA level exhibit a noticeable dependence on sample morphology. This indicates that morphology alone cannot account for the observed differences between the CG and AA models, as the atomistic samples were back-mapped from a parent CG trajectory and thus share the same morphology. Given that mechanical and elastic properties, as well as thermal transport dynamics, are strongly influenced by lattice dynamics \cite{hans2019_phonons}, it is reasonable to assume that the differences observed between the solid-state properties at the CG and AA levels are due to how lattice dynamics are represented in these models. The phonon density of states supports this assumption (see Fig. S7 in the supporting information), which shows a greatly simplified phonon spectrum in the CG samples, with only low-frequency modes (below 10 THz) represented. Although the MOLC model aims to provide physical properties in qualitative agreement with reference atomic simulations and experimental data \cite{roscioni2023osc}, the intrinsic reduction of vibrational degrees of freedom typical of mesoscopic models limits its sensitivity to the morphology of solid-state structures.

\section{Conclusions}

In this work, we tested and validated a new parametrization strategy for a coarse-grained (CG) model of PLGA polymers based on the MOLC force field, which was parametrized from the bottom-up using the OPLS-AA/M atomistic force field as a reference.

We investigated the amorphous and semi-crystalline phases of PLGA at the CG and all-atoms (AA) levels. For both phases, the CG model accurately reproduced the structural and thermal properties computed on the corresponding AA counterparts obtained through a back-mapping process. The density ($\rho$) of the CG models is in excellent agreement with the one from back-mapped AA simulations and the experimental value. The glass transition temperature ($T_g$) from CG simulations was found to depend on the sample morphology and to be higher than the experimental and reference values \cite{bellussi2022wettability}. The average error of 15\% on $T_g$ is attributed to the reduction of the degrees of freedom in the CG model of the polymer backbone, given that intermolecular interactions were calibrated to reproduce the cohesive energy computed at the AA level.

The thermal conductivity ($\lambda_c$) computed at the AA level was found to be in excellent agreement with the available experimental evidence, while the one computed at the CG level was found to be from 2 to 3 times smaller and not to vary with the sample morphology. 
Both the CG and AA models exhibited excellent agreement for the elastic properties. In particular, the computed values of Young's Modulus (E), Poisson's Ratio ($\nu$), and Tangential Modulus (G) were identical within the statistical uncertainty for the amorphous CG and AA models, while more significant differences between the two models were observed for the semi-crystalline morphology.

In conclusion, the CG model of PLGA exhibited good predictive capabilities for a broad range of solid-state properties, including density, glass transition temperature, thermal conductivity, and elastic constants. Some of those properties were affected by the loss of high-frequency vibrational modes in the CG model, resulting in a lack of sensitivity to the sample morphology.
In any case, these results demonstrate that the proposed CG model can replace the AA one while delivering a similar level of accuracy and a 10x enhancement in computational performance, paving the way for investigating larger and more complex polymer structures, which are crucial for the evaluation of surface properties. On top of that, the CG model allows to set up robust multiscale simulations where the accuracy of the sample can easily be increased by back-mapping to the atomistic structure, which only requires a short annealing to reach thermal equilibrium in the desired thermodynamical state.

A continuous improvement of materials models combined with a steady increase in computational power has made numerical simulations a reliable tool for predicting material properties and interpreting the results of complex experiments. These advances will ultimately enhance the speed of prototyping and reduce the time to market for developing new components. However, the difference in spatiotemporal scales between experiments and simulations requires resorting to continuum models, such as finite elements or finite volumes, to reach the scale of devices. Predicting the behaviour of materials at this scale is computationally prohibitive for most atomistic models. Instead, the CG model presented in this work is well-suited to model samples with dimensions comparable to those of the finite elements of continuum models, while delivering a nearly-atomic accuracy.

In this regard, it is worth noting that significant efforts have been dedicated over the past decade to developing discrete element methods (DEM), such as LIGGGTHS \cite{LIGGGTHS} or the atom to the continuum (ATC) \cite{ATC} package in LAMMPS, that in turn were used in hybrid MD-DEM simulations. We plan to use the MOLC CG model to evaluate the effects of controlled or uncontrolled surface roughness, as well as different phase morphologies, on the macroscopic wettability of PLGA surfaces \cite{provenzano2023experimentally,johansson1,johansson2}, closing the gap between mesoscopic and continuum models.

\begin{acknowledgement}

The authors thank CINECA (Iscra C and Iscra B projects) and the Politecnico di Torino's High Performance Computing Initiative (http://hpc.polito.it/) for the availability of computing resources. This project has received funding from the European Union’s Horizon research and innovation programme under grant agreement N. 760827 (OYSTER: Open characterisation and modelling environment to drive innovation in advanced nano-architectured and bio-inspired hard/soft interfaces).
We acknowledge PRACE for awarding us access to the Archer2 supercomputing facilities hosted by EPCC, Edinburgh (UK) under grant agreement N. 823767: SHAPE (SME HPC Adoption Programme in Europe), 14$^\text{th}$ programme. We sincerely thank Dr Gerhard Goldbeck for useful discussions on PLGA polymers.

\end{acknowledgement}

\begin{suppinfo}

The Supporting Information is available free of charge. Supporting Information covers additional details regarding the parametrization of PLGA and methods to evaluate mechanical and thermal properties.

\medskip
The optimized CG force field and reference AA force field for PLGA, and input scripts for MD simulations, are available free of charge in the Zenodo repository at \href{https://doi.org/10.5281/zenodo.13905472}{\url{10.5281/zenodo.13905472}}

\end{suppinfo}

\bibliography{molc}

\end{document}





\begin{figure}[t!]
	\centering
	\includegraphics[width=.95\textwidth]{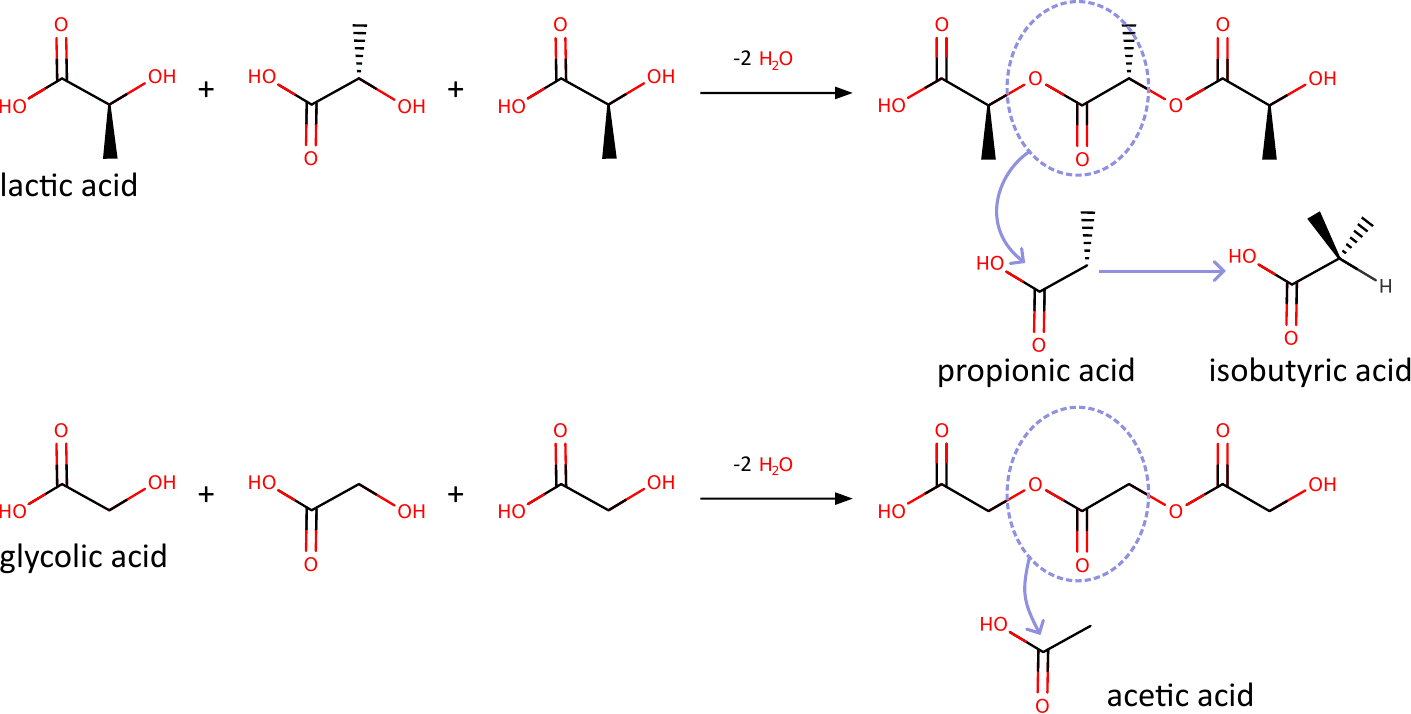}
	\caption[ ]{Skeletal formula of the condensation reaction yielding to Poly Lactic Acid -- PLA (top panel) and Poly Glycolic Acid -- PGA (bottom panel) polymers.
	As the condensation involves the elimination of a water molecule for every monomer added, the effective repeating unit of PLA is equivalent
	to propionic acid, while the repeating unit of PGA is equivalent to acetic acid, at least from a modelling perspective.
	The effective repeating unit of PLA was replaced with isobutyric acid to match the symmetry of the CG bead, given that the methyl group is found on alternated sides of the polymer backbone.
	The Gay-Berne parameters were fitted on Lennard-Jones potential curves computed from the OPLS-AA/M force field \cite{oplsaam}.
    }
	\label{PLGA-monomers}
\end{figure}

\begin{figure}[t!]
	\centering
	\includegraphics[width=.85\textwidth]{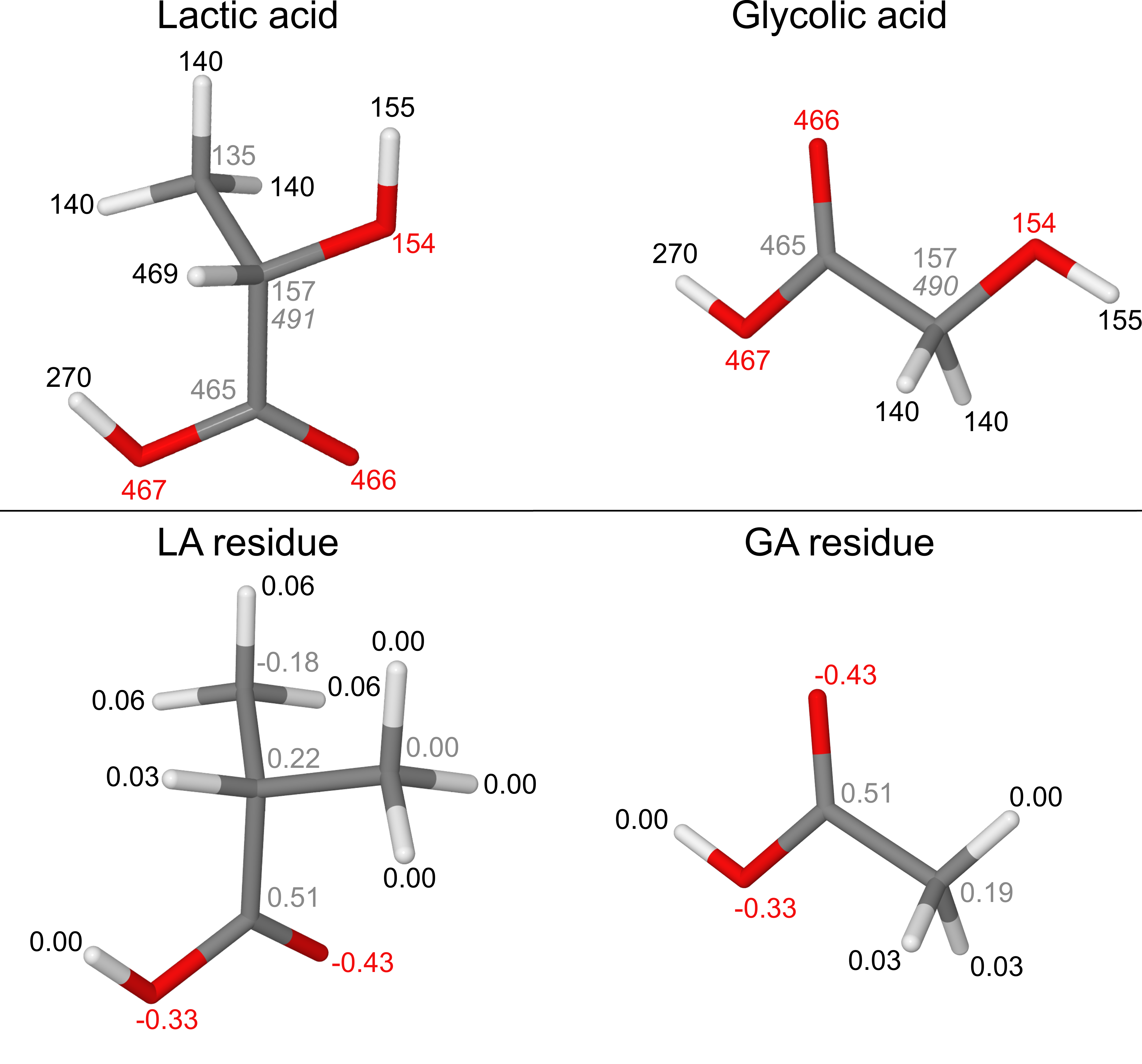}
	\caption[]{Top panel: OPLS-AA/M atom types for lactic and glycolic acid. The atom type for the carbon atom attached to the alcohol group is 157 for the isolated molecule and the terminal groups in PLGA polymers. The same carbon atom is changed to an ethyl ester with type 490 (\ce{CH2}) or 491 (\ce{CH}) for LA and GA residues in the PLGA polymer.\par
    Bottom panel: molecular structure and atomic charges of the effective residues used to model the repeating units of LA (isobutyric acid) and GA (acetic acid) in PLGA polymers, as described in Figure \ref{PLGA-monomers}.
	The charges are taken from the OPLS-AA/M force field \cite{oplsaam}, with the following modifications:
    \begin{itemize} 
	\item The charge of hydrogen atoms used to cap sigma bonds is set to zero.
	\item For the LA residue, the molecular structure is that of isobutyric acid, while the charges are that of propionic acid.
	\item The charges were rounded to make the repeating unit globally neutral. In this way, the terminal and central CG beads can use the same set of charges.
	\end{itemize}
	In addition, the Gay-Berne parameters were computed on the configurations shown in this picture.
	Colour code: carbon$/$grey, hydrogen$/$white, oxygen$/$red.}
	\label{PLGA-charges}
\end{figure}

\begin{figure}[t!]
	\centering
	\includegraphics[width=.95\textwidth]{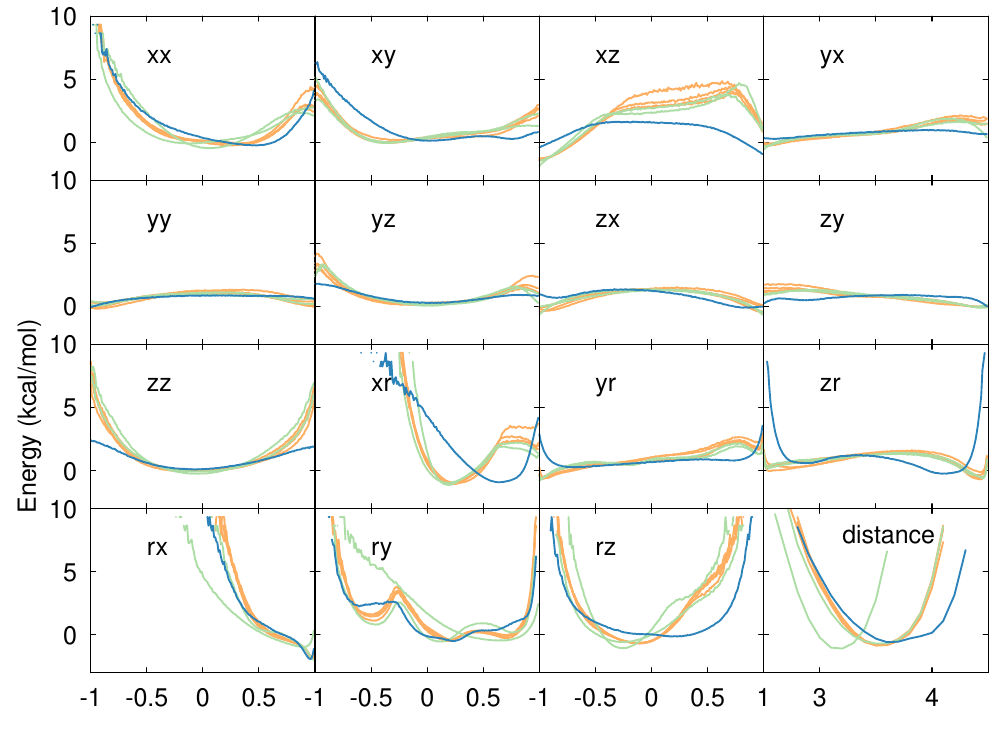}
	\caption[]{CG potential energy curves for two-body bonded interactions computed on LA-LA (orange), LA-GA (green), and GA-LA (blue) dimers. The panels $xx$ to $rz$ show the potential energy as a function of the scalar product between the ellipsoids' axes and the connecting vector $\overrightarrow{r}_{ij}$ (dimensionless), while the bottom-right panel shows the potential energy as a function of the scalar distance (\AA).
	The interaction potentials were obtained by Boltzmann inversion of the probability distribution functions computed from atomistic MD
	simulations. The AA to CG mapping is done by computing the inertia tensor of the atomistic residues, followed by the calculation of the
    scalar product between each component, plus the vector $\overrightarrow{r}_{ij}$ between the two particles. More details on the parametrization
    procedure can be found elsewhere \cite{molc}.}
	\label{PLGA-bonds}
\end{figure}

\begin{figure}[t!]
	\centering
	\includegraphics[width=0.95\textwidth]{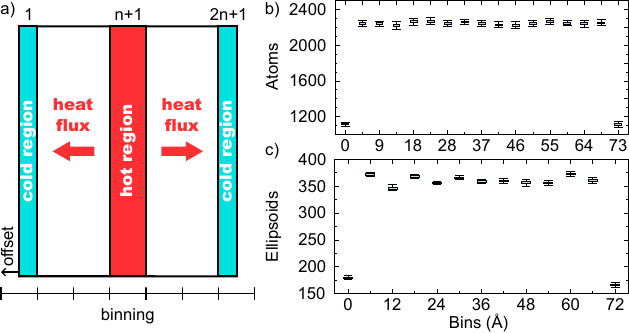}
	\caption[]{Panel a: setup employed to compute the thermal conductivity $\lambda_c$ from NEMD simulations. Each orthogonal cell vector is divided into $2n$ bins. An offset of half the bin length is used for binning the particles, resulting in a total number of $2n+1$ bins. The central bin is assigned to the hot region and the two lateral bins, connected through periodic boundary conditions, are assigned to the cold regions. Panels b and c: number of particles in each bin along the direction $a$ of a sample of amorphous PLGA. 17 bins ($n=8$) were used to compute the temperature profile of AA samples (b), while 13 bins ($n=6$) were used for the CG samples (c). Note that the number of particles in the lateral bins is approximately half that of a regular bin.}
	\label{binning}
\end{figure}

\begin{figure}[t!]
	\centering
	\includegraphics[width=.5\textwidth]{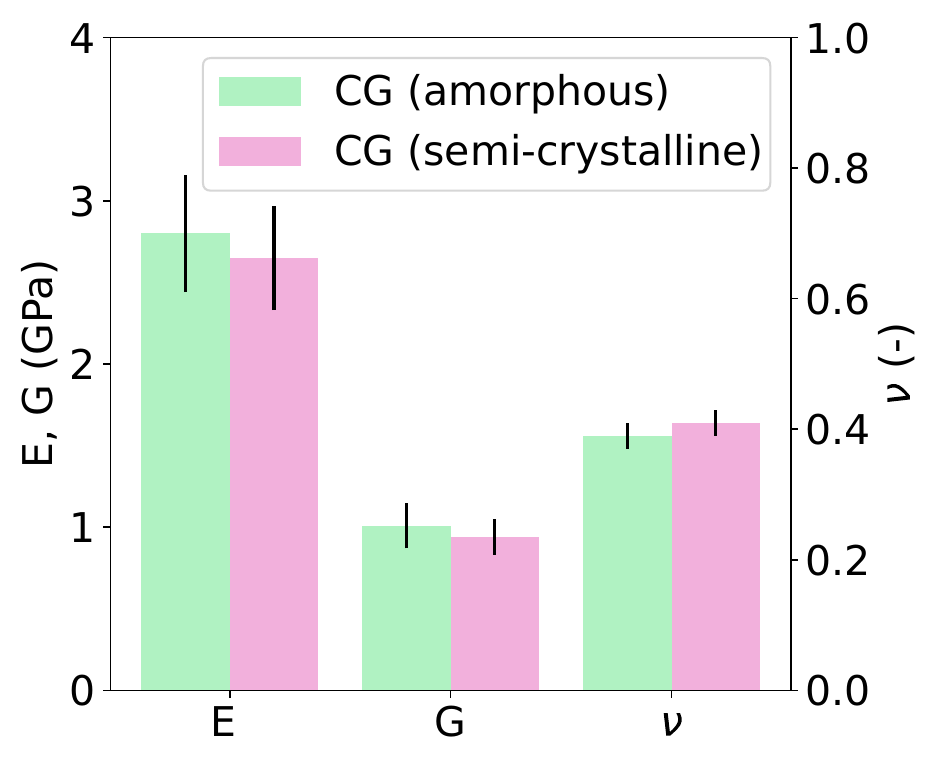}
	\caption{Comparison between the elastic ($E$) and shear ($G$) moduli, and Poisson ratio ($\nu$), computed for the amorphous (green) and  semi-crystalline (pink) morphologies, from CG-MD simulations.}
	\label{CG-elastic-comp}
\end{figure}

\begin{figure}[t!]
	\centering
    \newlength{\cgraph}\newlength{\ctext}
    \setlength{\ctext }{10em}
    \setlength{\cgraph}{.48\textwidth}
	\begin{minipage}[c]{\ctext} BM amorphous \end{minipage}%
	\begin{minipage}[c]{.725\cgraph}
		\includegraphics[width=\textwidth]{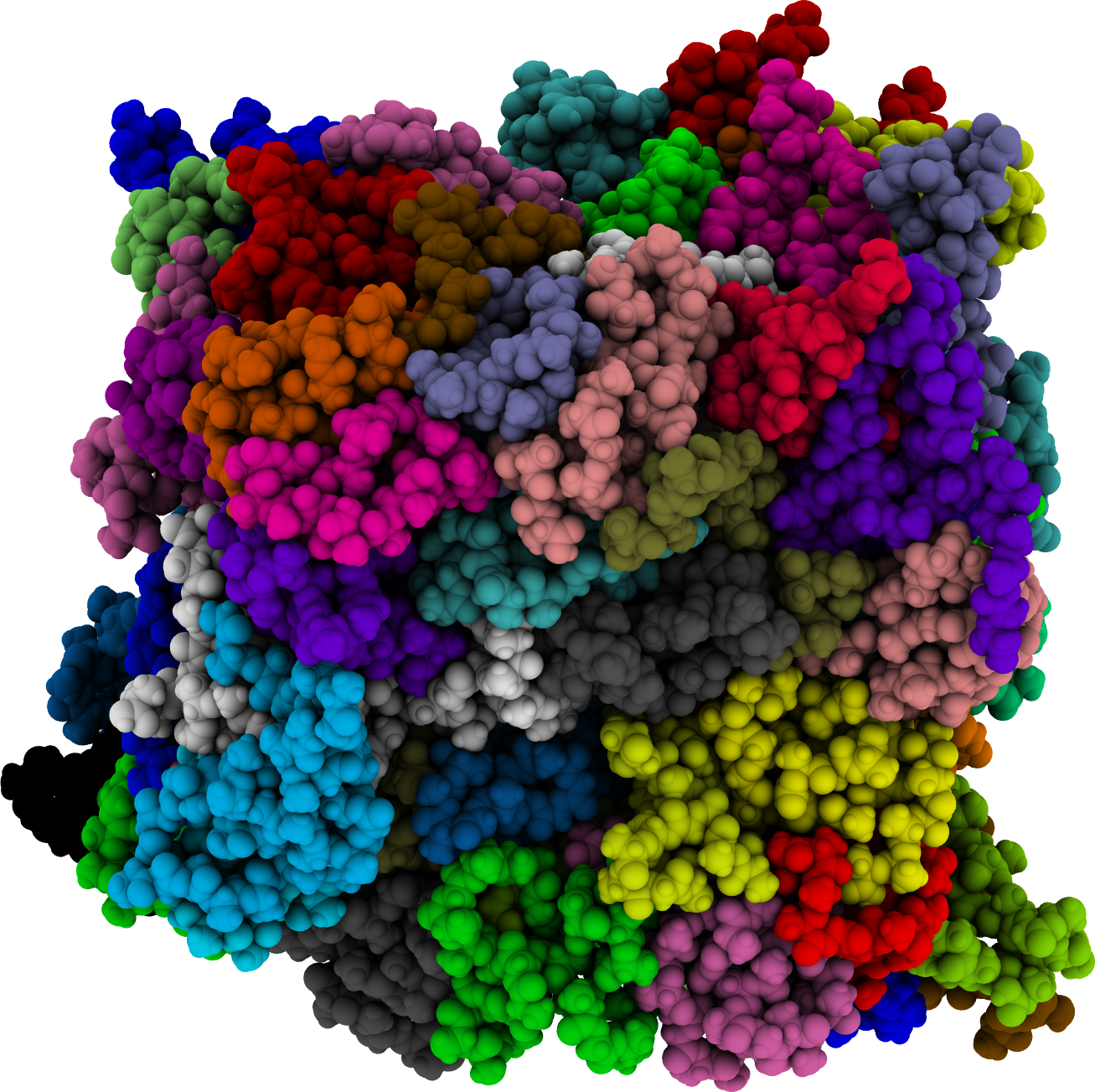}
	\end{minipage}
	\begin{minipage}[c]{\ctext} BM semi-crystalline \end{minipage}%
	\begin{minipage}[c]{\cgraph}
		\includegraphics[width=\textwidth]{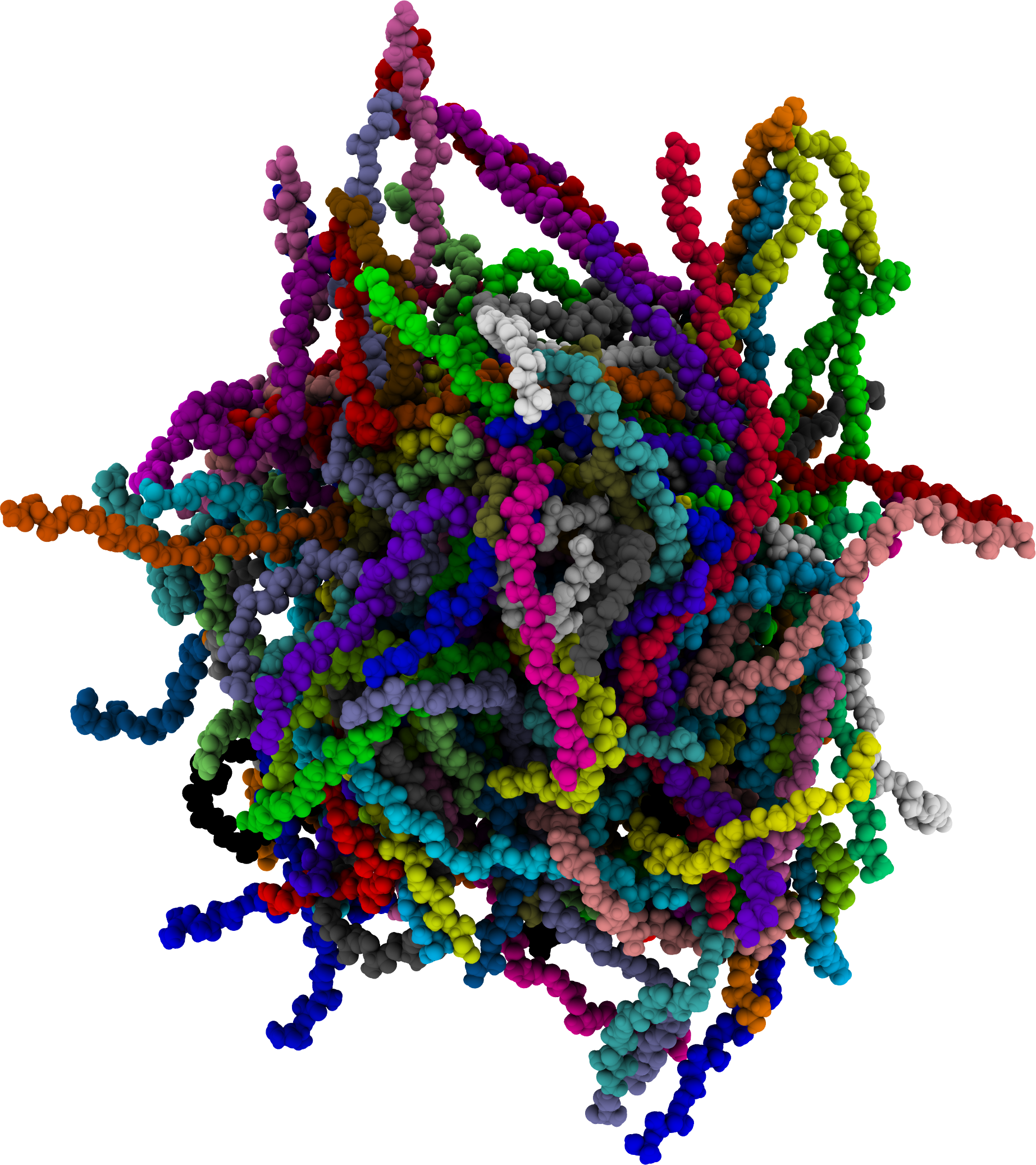}
	\end{minipage}
	\begin{minipage}[c]{\ctext}  AA amorphous \end{minipage}%
	\begin{minipage}[c]{.725\cgraph}
		\includegraphics[width=\textwidth]{figureS61}
	\end{minipage}
	\caption{Pictorial view of atomistic PLGA samples with individual chains coloured differently. The first two samples were obtained by back-mapping (BM) the corresponding CG configuration, while the last all-atom (AA) sample was created independently at the AA level and discussed in reference \cite{bellussi2022wettability}.}
	\label{plga-samples-aa}
\end{figure}

\begin{figure}[t!]
	\centering
	\includegraphics[width=\textwidth]{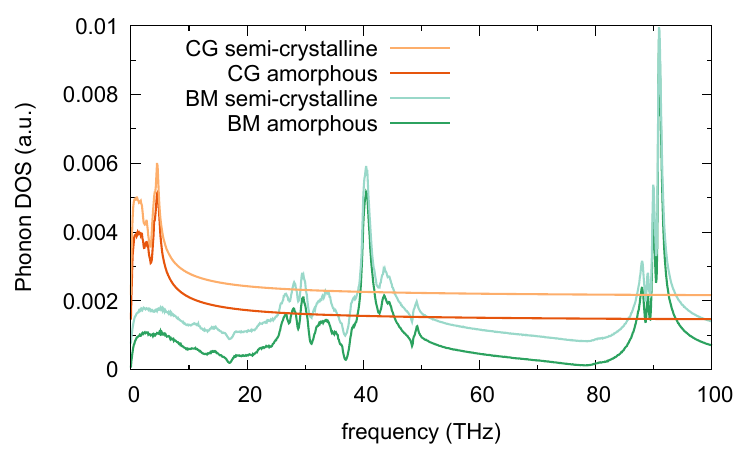}
	\caption[]{Phonon density of states (PDOS) computed for the CG and AA PLGA samples, with the AA samples obtained by back-mapping (BM) the corresponding CG configuration. The PDOS was obtained from the Fast Fourier transform (FFT) of the velocity--velocity autocorrelation function \cite{kong2011} $\langle v(t)v(0) \rangle$ computed from 100 consecutive MD trajectories, lasting 10 ps and having an integration timestep of 1 fs. In each trajectory, the signal is acquired at each integration time step, and the FFT is normalized to the length of the signal (\textit{i.e.} 10000 points over 10 ps). The PDOS show no dependence on the sample morphology.
    The PDOS of the CG samples differs from the AA ones due to the inherent simplifications introduced by the CG model. By grouping atoms into larger units, CG models average over detailed atomic interactions, leading to a loss of high-frequency vibrational modes associated with individual atomic motions. This reduction in degrees of freedom not only filters out short-wavelength modes but also alters the mass distribution and effective force constants within the system. The increased mass of the CG units shifts the vibrational modes toward lower frequencies, while the averaged interactions modify bond stiffness, affecting the overall vibrational spectrum. Consequently, the PDOS in CG model lacks the high-frequency contributions and detailed dynamic properties characteristic of atomistic models, as also noticed in previous works \cite{Muhammad2023,Yoo2024}.
 }
	\label{phonons}
\end{figure}

\clearpage
\bibliographystyle{achemso}
\bibliography{molc}